\documentclass[final,3p,times,twocolumn]{elsarticle}
 \biboptions{comma,sort&compress}
\usepackage{here}
 \usepackage{graphicx}
  \usepackage{epsfig}
\usepackage{comment}
\def\nin{\noindent}
\def\beq{\begin{equation}}
\def\eeq{\end{equation}}
\def\bea{\begin{eqnarray}}
\def\eea{\end{eqnarray}}

\def\tr{{\rm tr}}

\newcommand{\MeV}{\,{\rm MeV}}

\newcommand{\ignore}[1]{}
\journal{Nuc. Phys. (Proc. Suppl.)}

\begin{document}

\begin{frontmatter}



\title{Polyakov loop spectroscopy in the confined phase of gluodynamics and QCD}

 \author[label1]{E.~Meg\'{\i}as\corref{cor1}} \address[label1]{Grup de
   F\'{\i}sica Te\`orica and IFAE, Departament de F\'{\i}sica, Universitat
   Aut\`onoma de Barcelona, Bellaterra E-08193 Barcelona, Spain}
 \cortext[cor1]{Speaker} \ead{emegias@ifae.es}

 \author[label2]{E.~Ruiz Arriola} \address[label2]{Departamento de F{\'\i}sica
   At\'omica, Molecular y Nuclear and Instituto Carlos I de F{\'\i}sica
   Te\'orica y Computacional, Universidad de Granada, E-18071 Granada, Spain.}
\ead{earriola@ugr.es}

\author[label2]{L.L.~Salcedo}
\ead{salcedo@ugr.es}

\begin{abstract}
\noindent
By using a simple relativistic model, we compute the glueball and
gluelump spectra and relate these quantities, respectively, to the
trace anomaly and Polyakov loop in the adjoint representation of
gluodynamics. This spectroscopic description of thermodynamics is
extended with the inclusion of quarks. The relation between the hadron
resonance gas and the Polyakov loop in the fundamental and higher
representations is addressed.
\end{abstract}


\end{frontmatter}

% \linenumbers

\section{Introduction}
\label{sec:introduction}
\nin
%%%%%%%%%%%%
The confined phase of SU($N_c$) non-Abelian gauge theories can be naturally
described in terms of their relevant degrees of freedom in this regime. In
gluodynamics and QCD these are bound states of gluons and quarks,
i.e. glueballs and hadrons. By neglecting the interaction between these states
in the plasma, the hadron resonance gas model appears as a natural picture
that can describe the thermodynamics of these theories with a surprising
accuracy~\cite{Hagedorn:1984hz,Yukalov:1997jk,Agasian:2001bj,Tawfik:2004sw,Borsanyi:2010cj,Bazavov:2011nk,Megias:2012hk}.

The trace anomaly signals the breaking of scale invariance, and it is
a key quantity to study the thermodynamics of
QCD~\cite{Bazavov:2009zn}. Other quantity of interest is the Polyakov
loop (PL) in the fundamental representation, commonly used as an order
parameter for the confinement/deconfinement of color
charges~\cite{Svetitsky:1985ye}. While the interplay between the
PL and physical observables has been considered obscure for
a long time, recent advances have led to a description of the PL in terms of hadronic resonances, making a clear connection
between this quantity and the spectrum of
QCD~\cite{Megias:2012kb,Bazavov:2013yv,Megias:2013xaa}. In this
communication we will elaborate on this relation in gluodynamics and
QCD, and provide a physical interpretation of the PL in
representations other than the fundamental one.

%%%%%%%%%%%%
\section{Glueballs, gluelumps and thermodynamics}
\label{sec:gluodynamics}
\nin
%%%%%%%%%%%%%%%%%%%%%

Gluodynamics (and QCD) predicts the self-coupling of gluons. The
natural consequence is the possible existence of bound states with no
quarks, the so-called glueballs~\cite{Ochs:2013gi}. Paralleling
lattice studies (see e.g.~\cite{Lucini:2014paa}), some models have
been proposed to describe the lowest-lying
states~\cite{Brau:2004xw,Mathieu:2008pb,Buisseret:2009yv}. We will
study a simple relativistic model to obtain an overall description of
the glueball spectrum of two gluons. The model can be easily extended
to study the gluelumps: multigluonic states with one static gluon. We
will use the spectrum to compute the thermodynamics.

\subsection{Glueball Spectrum and Trace Anomaly}
\label{subsec:glueball}

The glueball is a bound state of two or more dynamical gluons. We consider a
relativistic model of two massless gluons. The Hamiltonian writes
\begin{equation}
H_{[gg]} = |\vec{p_1}| + |\vec{p_2}|  + V(r_{12}) \,,
\end{equation}
where the potential $V$ depends only on the relative distance between the
gluons. The classical partition function writes
\begin{equation}
\hspace{-0.6cm}
\log Z^{\textrm{\scriptsize cl}}_{[gg]} = 
\frac{\gamma^2}{2}
\int \frac{d^3x_1 d^3p_1}{(2\pi)^3} \frac{d^3x_2 d^3p_2}{(2\pi)^3}
e^{-H_{[gg]}/T} + \cdots ,
\label{eq:logZ}
\end{equation}
where $\gamma$ is the gluon spin degeneracy and the dots stand for 3
and higher gluon terms which will be neglected.  In what follows we
assume a gluon-gluon potential of the form $V(r) = \sigma_8 r$, with
Casimir scaling between the adjoint and fundamental representation
string tensions $\sigma_8 = \frac{N_c^2}{4}\sigma_3$. The trace
anomaly is obtained from the standard thermodynamic relation
\begin{equation}
\Delta(T) \equiv T\frac{\partial}{\partial
  T}\left( \frac{P}{T^4} \right) = \frac{\varepsilon - 3 P}{T^4}  \,,
\quad
Z= e^{PV/T} . \label{eq:DeltaT}
\end{equation}
After an explicit evaluation of Eq.~(\ref{eq:logZ}),
the classical result for the trace anomaly in gluodynamics is
\begin{equation}
\Delta^{\textrm{\scriptsize cl}}_{[gg]}(T) 
= \frac{\gamma^2}{2} \frac{48}{\pi^3 \sigma_8^3} T^6 \,.
\label{eq:Delta0}
\end{equation}

\begin{figure}[t]
\begin{center}
\epsfig{figure=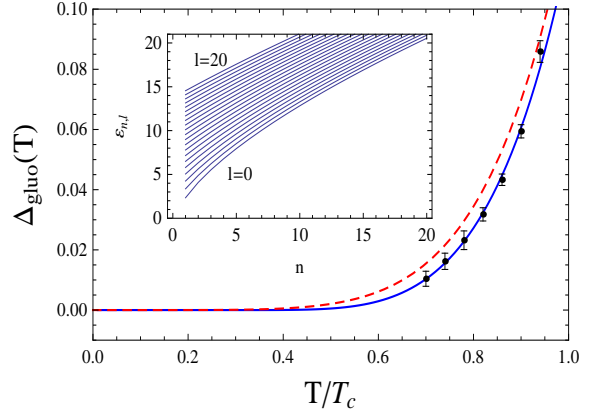,height=55mm,width=75mm}
\end{center}
\vspace{-0.5cm}
\caption{Trace anomaly of gluodynamics as a function of temperature
  (in units of~$T_c$).  We plot as continuous (blue) line the quantum
  result from Eqs.~(\ref{eq:logZquantum})-(\ref{eq:Schr}), and as
  dashed (red) line the classical result, Eq.~(\ref{eq:Delta0}). We
  use $\gamma=3$. Lattice data have been taken from
  \cite{Borsanyi:2012ve}. The inserted figure corresponds to the
  spectrum obtained with Eq.~(\ref{eq:Schr}).}
\label{fig:tracequantumsemi}
\end{figure}

In order to isolate the relative motion and deal with a simpler 
quantum Hamiltonian, it is convenient to transform (\ref{eq:logZ}) by
applying the identity~\cite{Antonov:2006bv}
\begin{equation}
e^{-|\vec{p}|/T} = \frac{1}{\sqrt{\pi T}}
\int_0^\infty \frac{d\mu}{\sqrt{ \mu}}
e^{- \frac{1}{T} \, \left(\frac{\vec{p}^2}{4\mu} + \mu \right)  } 
\end{equation} 
to both gluons. Integrating out the center of mass gives
\begin{equation}
\hspace{-7mm}
\frac{\log Z^{\textrm{\scriptsize cl}}_{[gg]}}{V} \!=\!  \frac{4
  \sqrt{T}}{3\pi^{5/2}} 
\!\!\int_0^\infty \!\!\!\!\!d\mu \mu^{3/2} e^{-\frac{\mu}{T}} 
\frac{\gamma^2}{2}
\!\! \int \!\!\frac{d^3x d^3p}{(2\pi)^3} 
e^{-\frac{1}{T}\left( \frac{\vec{p}^2}{\mu} + V(r)\right)}
\label{eq:logZ2}
\end{equation}
This auxiliary system can now be quantized by applying standard
quantization rules.\footnote{Note that this is not identical to a
  direct quantization of $H_{[gg]}$, which is technically much
  harder.} Then, Eq.~(\ref{eq:logZ2}) transforms into the quantum
partition function
\begin{equation}
\hspace{-0.7cm}\frac{\log Z_{[gg]}}{V} \! = \!\frac{4\sqrt{T}}{3\pi^{5/2}}
\!\!\int_0^\infty \!\!\!\!d\mu \mu^{3/2} e^{-\frac{\mu}{T}} \!\!\sum_{n,l}
\nu_{l}(2l+1)e^{-\frac{1}{T}\frac{\sigma_8^{2/3}}{\mu^{1/3}}\varepsilon_{n,l}} ,
\label{eq:logZquantum}
\end{equation}
where $\nu_{l}=\gamma(\gamma\pm1)/2$ for even/odd $l$ and the
spectrum is readily obtained from the Schr\"odinger equation
\begin{equation}
\left(-\nabla^2 + r\right) \Psi_{n,l} = \varepsilon_{n,l} \Psi_{n,l} \,.
\label{eq:Schr}
\end{equation}
We plot in Fig.~\ref{fig:tracequantumsemi} the result of the trace anomaly by
using Eqs.~(\ref{eq:logZquantum})-(\ref{eq:Schr}).
We have fitted the lattice data by considering as the sole fitting
parameter the dimensionless ratio $T_c/\sqrt{\sigma_3}$. The result is
$T_c/\sqrt{\sigma_3} = 0.774$ and $\chi^2/\textrm{dof}=0.54$, in the
regime $ 0.70 < T/T_c < 0.94$ with $\gamma=3$. It is noteworthy that
such a simple model leads to a value which is similar to the one
reported from lattice computations $T_c/\sqrt{\sigma_3} \approx
0.629(3)$~\cite{Borsanyi:2012ve}. The agreement between classical and
quantum results for temperatures $T \gtrsim 0.7 T_c$ can be improved
by including the first semiclassical correction.

\subsection{Gluelump spectrum and Polyakov loop}
\label{subsec:gluelump}

The similar model for the gluelump spectrum (an adjoint source dressed by
dynamical gluons) in the leading approximation (just one gluon) has the
Hamiltonian
\begin{equation}
H_{[Gg]} = |\vec{p}| + V(r) \,.
\end{equation}
Simple scaling shows that $M_{Gg}=M_g/\sqrt{2}$ for the ground states of
glueball and gluelump in the $\sigma_8 r$ model. A remarkable consequence is
that, strictly speaking, the smallest mass gap in gluodynamics is not the
lightest glueball mass but the lightest gluelump mass.

Straightforward integration in the equation similar to~(\ref{eq:logZ})
gives the explicit result for the {\it classical} partition function
\begin{equation}
L_{\bf 8}(T) \approx \log Z^{\textrm{\scriptsize cl}}_{[Gg]} 
= \gamma \frac{8}{\pi\sigma_8^3} T^6 \,, 
\label{eq:logZgluelumpalphas}
\end{equation}
where $L_{\bf 8}$ is the adjoint PL~\cite{Megias:2013xaa}. From a comparison of
Eqs.~(\ref{eq:Delta0}) and (\ref{eq:logZgluelumpalphas}), one gets the
approximate scaling~$L_{\bf 8}(T) \approx \frac{\pi^2}{3\gamma} \Delta_{[gg]}(T)$.

The {\it quantum} version, similar to (\ref{eq:logZquantum}), for the
partition function of the gluelump is
\begin{equation}
\hspace{-0.5cm}\log Z_{[Gg]} = \gamma \int_0^\infty
\!\!\!\!\frac{d\mu}{\sqrt{\pi T \mu}} e^{-\frac{\mu}{T}} \!\!\sum_{n,l}
(2l+1)e^{-\left(\frac{\sigma_8^2}{4\mu}\right)^{1/3}\frac{\varepsilon_{n,l}}{T}} , 
\label{eq:logZgluelumpquantum}
\end{equation}
with the same eigenvalues $\varepsilon_{n,l}$ as in Eq.~(\ref{eq:Schr}).

\begin{figure}[t]
\begin{center}
\epsfig{figure=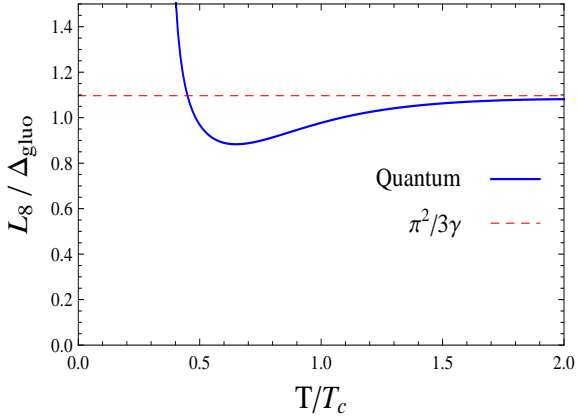,height=55mm,width=75mm}
\end{center}
\vspace{-0.5cm}
\caption{Ratio between PL in the adjoint representation and trace
  anomaly, as a function of temperature (in units of $T_c$). Continuous (blue)
  line is the quantum result from
  Eqs.~(\ref{eq:logZquantum})-(\ref{eq:Schr}) and
  (\ref{eq:logZgluelumpquantum}). Dashed (red) line is the approximate
  classical scaling $\pi^2/3\gamma$. We use $\gamma=3$. 
}
\label{fig:traceL8}
\end{figure}

The ratio between the adjoint PL and the trace anomaly computed
with the model is plotted in Fig.~\ref{fig:traceL8}. The approximate classical
scaling $\pi^2/3\gamma$ is valid for temperatures $T \gtrsim 0.5 T_c$.

%%%%%%%%%%%%
\section{QCD thermodynamics and Polyakov loop}
\label{sec:hrgm}
\nin
%%%%%%%%%%%%%%%%%%%%%

An effective approach to QCD at finite temperature is provided by
chiral quark models coupled to the Polyakov
loop~\cite{Fukushima:2003fw,Megias:2004hj,Megias:2006bn,Ratti:2005jh,Dutra:2013lya,Ferreira:2013tba}. While
the mean field approximation is widely used to study the phase
transition, it was stressed
in~\cite{Megias:2004hj,Megias:2005qf,Megias:2006bn,Megias:2006df,Megias:2006ke}
that local and quantum corrections of the PL are needed at low
temperatures. We will use this approach to derive the relation between
the QCD spectrum and the PL in fundamental and higher representations.

\subsection{Polyakov-Constituent Quark Model}

The partition function of the model
is~\cite{Megias:2004hj,Sasaki:2012bi,Megias:2013aua,Megias:2013xda,Megias:2013xaa}
\begin{equation}
Z = \int {\cal D}\Omega {\cal D}\bar{q} {\cal D}q \, 
e^{-S_g (T,\Omega) - S_q (T,\Omega)} \,,
\end{equation}
where the gluonic and quark actions read
\begin{eqnarray}
&&\hspace{-1.4cm}S_g =  -2 T \!\!\int \!\!\frac{d^3 x d^3 p}{(2\pi)^3} \tr \log \left(1 \!-\! \Omega_8(\vec{x})  \, e^{-E_p/T}  \right) \,, \\
&&\hspace{-1.4cm}S_q =  2 N_f \!\! \int \!\!\frac{d^3x d^3p}{(2\pi)^3} \bigg( 
 \tr \log \big[ 1\!+\!\Omega_3 (\vec{x}) \, e^{-E_p/T} \big] \!+\!h.c.  \bigg) \,.
\end{eqnarray}
The PL variables $\Omega_{\mu}(\vec{x})$ and
$\Omega_{\mu}^\dagger(\vec{x})$ play the role of quark and gluon
creation/annihilation operators. A series expansion of the action in
powers of these operators leads to an expansion in the number of
constituents, and each term can be identified with a
multi-quark(gluon) state.  In particular the meson contribution to the
partition function follows from the correlator $Z_{[q\bar{q}]}
\!\sim\! \langle \tr\,\Omega_3(x) \tr\,\Omega_3^\dagger(x^\prime)
\rangle \!\sim\! e^{-(M_q + M_{\bar{q}})/T}$. After quantization as in
Sec.~\ref{sec:gluodynamics}, the Boltzmann factor at low temperature
will contain the spectrum of mesons. There are also contributions from
baryons, glueballs, etc. If we neglect the interaction between hadrons
and retain only the confining forces, one has $Z \simeq Z_{[q\bar{q}]}
Z_{[qqq]} Z_{[gg]} \cdots$. This approximation, known as the Hadron
Resonance Gas (HRG) model, has been widely used to describe the confined
phase of QCD with a remarkable agreement to lattice
data~\cite{Bazavov:2009zn,Megias:2012hk}.

\subsection{Hadron Resonance Gas model and Polyakov loop}

The partition function described above corresponds to a plasma formed by
dynamical constituents. An alternative physical system appears when one
considers in the plasma a static color source (heavy quark) in position $x_0$
and representation ${\bf 3}$. This source polarizes the medium, as it becomes
screened by dynamical quarks and gluons to form a heavy-light hadron which is
color neutral. The consequence is that the partition function of this system
receives contributions of the form
\begin{eqnarray}
&&\hspace{-1.4cm}Z_{[Q\bar{q}]} \!\sim\! \langle \tr\,\Omega_3(x_0) \,\tr\,\Omega_3^\dagger(x^\prime) \rangle \!\sim\! e^{-(M_Q + M_{\bar{q}})/T} \,,  \\
&&\hspace{-1.4cm}Z_{[Qqq]} \!\!\sim\!\! \langle \tr\, \Omega_3(x_0) \,\tr\,\Omega(x^\prime)  \,\tr\,\Omega(x^{\prime\prime}) \rangle \!\sim\! e^{-(M_Q + 2M_q)/T} \,,  \\
&&\hspace{-1.4cm}Z_{[Q\bar{q}g]} \!\!\sim\! \langle \tr\Omega_3(x_0) \, \tr\Omega_3^\dagger(x^\prime) \, \tr \Omega_8(x^{\prime\prime})  \rangle \!\sim\!\! e^{-(M_Q+M_{\bar{q}}+M_g)/T}
\end{eqnarray}
After quantization and renormalization~\cite{RuizArriola:2012wd}, one gets the
HRG model for the PL in the representation ${\bf 3}$, which can be
written as
\begin{equation}
\hspace{-0.5cm}L_{\bf 3} \simeq Z_{[Q\bar{q}]} Z_{[Qqq]} Z_{[Q\bar{q}g]} \cdots  \simeq \frac{1}{2N_c} \sum_{[Q\alpha]} g_{[Q\alpha]} e^{-\Delta_{[Q\alpha]} /T} \,, \label{eq:hrgmL3}
\end{equation}
where $\Delta_{[Q\alpha]} = \lim_{m_Q \to \infty} (M_{[Q\alpha]} - m_Q)$. In
these formulas $[Q\alpha]$ stands for the spectrum of heavy-light hadrons, and
$m_Q$ is the mass of the heavy quark. The fact that the PL
corresponds to a partition function explains its real and positive
character. This approach was first proposed in~\cite{Megias:2012kb}, and it
describes very well the lattice data for the PL in the fundamental
representation in the confined phase when a large enough amount of states are
included in the spectrum~\cite{Megias:2012kb,Bazavov:2013yv}.

\begin{figure}[t]
\begin{center}
\epsfig{figure=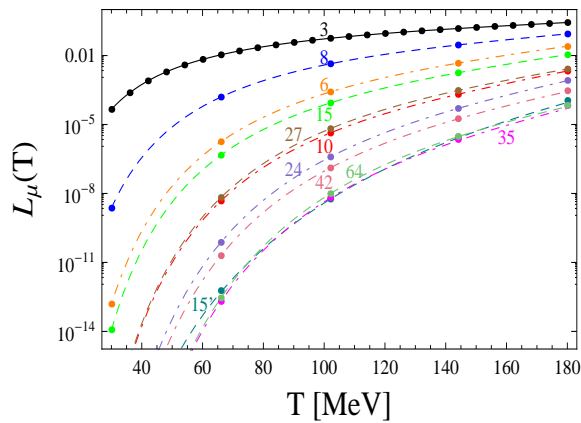,height=55mm,width=75mm}
\end{center}
\vspace{-0.5cm}
\caption{Polyakov loop in several representations as a function of $T$ (in
  $\MeV$). From top to bottom $\mu= {\bf 3}$, ${\bf 8}$, ${\bf 6}$, ${\bf
    15}$, ${\bf 27}$, ${\bf 10}$, ${\bf 24}$, ${\bf 42}$, ${\bf 64}$, ${\bf
    15'}$, and ${\bf 35}$. We consider $N_f=2$, $M_q=300\,\MeV$ and
  $M_g=664\,\MeV$.}
\label{fig:Lmu}
\end{figure}

It is possible to generalize this model to the PL in any
representation~$\mu$. In this case $Q$ is replaced by a static source in the
representation~$\mu$, which we denote by $S_\mu$, and it is screened by
dynamical constituents to form color neutral states according to a specific
pattern which depends on $\mu$~\cite{Megias:2013xaa}. The result is
\begin{equation}
\hspace{-0.5cm}L_{\mu} \simeq \frac{1}{2N_c} \sum_{[S_\mu\alpha]} g_{[S_\mu\alpha]} e^{-\Delta_{[S_\mu\alpha]} /T} \,. \label{eq:hrgmLmu}
\end{equation}
We show in Fig.~\ref{fig:Lmu} the PL in several representations
computed with the Polyakov-Constituent Quark model.
The low temperature behavior is dominated by the lightest energy gap
associated with the screening of heavy sources, and as such, implements
particular scaling rules between the PL in different
representations, which in general are different from the Casimir
scaling~\cite{Gupta:2007ax}. A list of alternative low temperature scaling
rules has been identified in~\cite{Megias:2013xaa}, and could be tested by
lattice calculations.

%%%%%%%%%%%%%%%%
\section{Discussion and outlook}
\label{sec:conclusions}
\nin
%%%%%%%%%%%%%%%%

The previous considerations provide interesting guidelines to model
not only the trace anomaly but also the interesting physical situation
appearing when a static color source is placed in the hot but confined
medium. In this case the medium becomes polarized in the color degrees
of freedom, as dynamical colored particles tend to screen the
source. The partition function of this system is related to the
expectation value of the PL in the group representation of the static
source and admits a hadronic representation in terms of bound states
in which the source appears as one of the constituents.  The natural
extension to QCD allows the static source to be either a quark, gluon
or combination of both, and the spectrum is formed by conventional
heavy-light hadrons, and possibly hybrid and exotic states. From this
point of view, the possibility of using the Polyakov loop in higher
representations than the fundamental one, emerges as a fascinating
opportunity to study the spectroscopy of QCD, including excited
states, exotics and hybrids which could be tested on the lattice.

%%%%%%%%%%%%%%%%%%%%%%%%%%%
\section*{Acknowledgements}
\nin
%%%%%%%%%%%%%%%%
This work has been supported by DGI (FIS2011-24149 and FPA2011-25948), Junta
de Andaluc{\'\i}a grant FQM-225, Spanish MINECO Consolider-Ingenio 2010
Program CPAN (CSD2007-00042) and Centro de Excelencia Severo Ochoa Programme
grant SEV-2012-0234. The research of E.M. is supported by the Juan de la
Cierva Program of the Spanish MINECO.


%% References


%\bibliographystyle{elsarticle-num}
%\bibliography{refs}


%%%%%%%%%%%%%%%%%%%%
%\vfill\eject

%\input{bib_sample}

%\bibliographystyle{h-physrev3}
%\bibliography{Refs}

\end{document}